\documentclass[a4paper,12pt]{report}
\usepackage[utf8]{inputenc}
\usepackage{graphicx}
\usepackage{subfigure}
\usepackage{hyperref}
\usepackage{cleveref}
\usepackage{cite}
\usepackage{setspace}
\usepackage[nottoc,notlof,notlot]{tocbibind} 
\usepackage{titling}
\preauthor{\begin{flushleft}\large}
\postauthor{\par\end{flushleft}}

\begin{document}
\doublespacing
\author{\normalsize Nadeen R. Rishani, Hadeel Elayan, Raed M. Shubair, and Asimina Kiourti}
\title{Wearable, Epidermal and Implantable Sensors for
Medical Applications}
\date{2018}

%\frontmatter
\maketitle
\tableofcontents

%\mainmatter
\chapter{Internet of NanoThings}
%\begin{abstract}
Continuous health monitoring using wireless body area networks (WBANs) of wearable, epidermal and implantable medical devices is envisioned as a transformative approach to healthcare. Rapid advances in biomedical sensors, low-power electronics, and wireless communications have brought this vision to the verge of reality. However, key challenges still remain to be addressed.  This paper surveys the current state-of-the-art in the area of wireless sensors for medical applications. Specifically, it focuses on presenting the recent advancements in wearable, epidermal and implantable technologies, and discusses reported ways of powering up such sensors. Furthermore, this paper addresses the challenges that exist in the various Open Systems Interconnection (OSI) layers and illustrates future research areas concerning the utilization of wireless sensors in healthcare applications.
%\end{abstract}


% \fi

\section{Introduction}
The last two decades have witnessed an exponential growth and tremendous developments in wireless technologies and systems, and their associated applications, such as those reported in \cite{shubair_robust_2005,shubair_performance_2005, belhoul_modelling_2003, shubair_robust_2004, al-ardi_direction_2006, nwalozie_simple_2013, al-nuaimi_direction_2005, bakhar_eigen_2009, alhajri2018accurate, samhan2006design, alhajri2015hybrid, mohjazi2011deployment, shubair_convergence_2005, shubair2007setup, al2005direction, shubair_improved_2006, shubair_displaced_2007, hakam_accurate_2013, shubair_enhanced_2015, shubair_robust_2006, shubair_enhanced_2009, shubair_detection_2011, alkaf_improved_2012,hakam_robust_2014-1, shubair_performance_2006, alayyan_mmse_2009, shubair_improved_2007, jimaa_performance_2008, kulaib_investigation_2013, kulaib_improved_2015, kulaib_accurate_2015, alhajri_hybrid_2015-1, goian_fast_2015, shubair_simple_1992, kulaib_performance_2011, shubair2013adaptive, hakam_novel_2016, moghaddam_novel_2010, lazovic_comparative_2013, samahi_performance_2006, ardi_adaptive_2004, jimaa_convergence_2009, al-ardi_performance_2003-1, shubair_displaced_2008, hakam_enhanced_2013, hakam_robust_2014, shubair_improved_2005, al-ardi_performance_2003, samhan_design_2006, al2003investigation, al-ardi_computationally_2005, al-ardi_computationally_2004, kulaib_overview_2011, shubair2015vivo,khan_compact_2017, shubair_closed-form_1993, omar_uwb_2016, elayan2017terahertz, elayan2017wireless, elayan2016channel, elayan2016vivo, elayan2017bio, elayan2018end, elayan2017photothermal, elayan2017multi, elayan2018vivo, alnabooda2017terahertz, elayan_towards_2018}.Wireless Body Area Networks (WBANs) are a new generation of Wireless Sensor Networks (WSNs) dedicated for healthcare monitoring applications. The aim of these applications is to ensure continuous monitoring of the patients' vital parameters, while giving them the freedom of moving. In doing so,  WBANs result in an enhanced quality of healthcare \cite{shubair2015vivo}. Advanced health care delivery  relies on both body surface (wearable) and internal sensors (implants) \cite{ketterl2012vivo}. The benefit provided by WBAN is obvious to the patient's comfort especially for long-term monitoring as well as complex monitoring during surgery and medical examinations \cite{wegmuller2007intra}.

Nonetheless, meeting the potential of WSNs in healthcare necessitates addressing a number of technical challenges \cite{hadeel}. These challenges reach beyond the resource limitations that all WSNs face in terms of limited network capacity, processing and memory constraints, as well as scarce energy reserves. Specifically, unlike applications in other domains, healthcare applications impose stringent requirements on system reliability, quality of service as well as privacy and security \cite{ko2010wireless}. 

In this review paper, we describe the current state of research and development of wireless sensors for medical applications. The most recent developments in terms of wearable,epidermal and implantable devices are  presented. In addition, we expand on both the challenges and future directions associated with wireless sensors for healthcare.  The rest of the paper is organized as follows. In Section~II, we present the requirements of wireless sensors for medical applications.  In Section~III, Section~IV and Section~V, the current features and  recent advances of wearable, epidermal and implantable wireless body networks are discussed, respectively. In Section~VI, powering considerations which entail a major metric in medical healthcare, is described.  Challenges along with future research directions in wireless body area networks are illustrated in Section~VII and Section~VIII. Finally, we draw our conclusions in Section~IX.
\section{Requirements for Wireless Medical Sensors}
In order to sense biological information from either outside or inside the human body, wearable and implantable sensors are, respectively, utilized. These sensors typically communicate the acquired information to a control device worn on the body or placed in an accessible location. Subsequently, the data assembled from the aforementioned control devices are conveyed to remote destinations in a WBAN for diagnostic and therapeutic purposes. To do so, wireless networks with long-range transmission capabilities need to be integrated \cite{6831489}. Specifically,  sensors used in wireless networks for healthcare applications must satisfy the following requirements:
\vspace{-7 pt} 
\subsection{Unobtrusiveness}\label{battery}
The most essential requirement in the design of wireless medical sensors relates to their light weight and miniature size, for these characteristics allow both non-invasive and unobtrusive continuous monitoring of health \cite{7529464}. The size and weight of a sensor mostly rely on the size and weight of its battery, where a battery's capacity is directly proportional to its size. Recent technological advances in microelectronics, system-on-chip design and low power wireless communication led to the development of small size, high energy batteries \cite{li2010magnetoelectric}. Flexible and printed batteries both hold promise for wearable devices. Multiple wearable applications exist where flexible batteries offer a distinct advantage, including  skin patches for transdermal drug delivery, patient temperature sensors, or RFID tracking \cite{7303932}.
\vspace{-7 pt} 
\subsection{Security}
One of the essential design fundamentals of a WBAN is the security of the entire system. That is, data integrity must be ensured, which implies that sensors must fulfill the privacy requirements provided by law. The utmost goal is to enable secure and efficient wireless networks where data are readily  accessible by any authorized person, even in remote destinations. Coordination between the system hardware and related security software components is fundamental to providing secure and reliable communication \cite{7042603}.
\vspace{-7 pt} 
\subsection{Interoperability}
Interoperability in healthcare is the extent to which various systems and devices can  interpret  data and display it in a user-friendly way. This entails that  data exchange methods will allow information to be shared across hospitals, pharmacies, labs, clinicians and patients, regardless of which vendor is used. The main objective behind interoperability is to reform the chaotic and at times dysfunctional nature of  information exchange among hospitals. Through interoperability, data becomes exceptionally mobile. Personal health information, entered into a system once, becomes available to patients wherever they are and whenever they need it \cite{7503819}. 
\vspace{-7 pt} 
\subsection{Reliable Communication}
For medical applications that rely on WBANs, the reliability of the communication link is of paramount importance. The communication constraint varies between nodes since the sampling rates required by each sensor are different. For example, instead of sending raw electrocardiogram (ECG) data from sensors, we can perform feature extraction on the sensor, and transfer only information about the particular event. In addition to reducing the high demands on the communication channel, the reduced communication requirements save total energy expenditures, and consequently increase battery life. A careful trade-off between communication and computation is crucial for optimal system design \cite{6974442}.

\section{Wearable Wireless Body Area Networks}\label{sec3}
In order to carefully track discrepancies  in patients' vital activities and provide feedback for maintaining optimal health status,  health monitoring systems have been introduced. These systems can be in the form of either devices placed on the body (e.g. bracelets, watches, etc) or devices based on electronic textiles incorporated into fabric.
Upon their integration into the telemedicine system, the wearables turn into alert systems notifying the medical staff when life-threatening changes occur within the patient's body \cite{lymberis2004wearable}. Long-term continuous monitoring can as well be attained as part of the diagnostic procedure.  Such monitoring may confirm adherence to treatment guidelines and help assist the effects of drug therapy. 

For example, heartbeat and respiration rate recording was implemented in \cite{provisional} using a wearable device. The system consists of a highly sensitive tri-axial accelerometer, a temperature sensor, an air pressure sensor as well as a central node that transfers measured data to a PC or handheld device via wireless communication in the ISM band.
Moreover, ``NASA'' is developing a wearable patch that controls heart rate, blood pressure and other physiological parameters for astronauts \cite{lin2000wearable}. In addition, the ``Nike-Apple'' iPod Sports kit as well as the ``Lifeshirt'' developed by ``In Vivo Metrics'' are amongst some of the most common commercial wearable prototypes \cite{diamond2008wireless}. As diabetes continues to be a major health issue, the authors in \cite{insulin}  used wearable devices for blood glucose control. The device is composed of a sensor that collects information about blood glucose, and an insulin automatic injection pump coupled with an Enhanced Dynamic Closed-loop Control algorithm (ED- CLC).

Furthermore, advances in smart technologies gave rise to a wearable industry which is a key enabler of optimal progressions in our societies. For instance, ``Smart Stop'' by Chrono Therapeutics, is a smart device that aims to help people stop smoking. The device is embedded with sensors that sense changes in the body and put into motion algorithms that detect the  craving of a person for  cigarette and nicotine. In turn, the device delivers medication to the person so that the craving can be curtailed.  Another example includes ``Google Smart Contact Lenses'' \cite{8000289}. Basically, Google's smart contact lenses  are made for people who suffer either from diabetes or for those who simply wear glasses. The technology is engineered to take the tears in a person's eye and measure the glucose levels that are present. For people who wear glasses, the lens would be engineered to restore the eye's natural autofocus.

\begin{figure}[!t]
\centering{\includegraphics[angle=-90,width=0.4\textwidth]{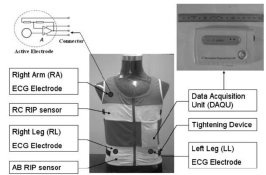}}
\caption{T-shirt with sensors integrated for wearable cardiopulmonary parameter monitoring \cite{chinashirt}.
%\figfooter{a}{Subfigure 1}
%\figfooter{b}{Subfigure 2}
\label{china}}
\end{figure}

Textile-based devices for medical applications are integrated into fabrics including patient's cloths or blankets, all of which facilitate wireless health monitoring \cite{nadeenchapter}.  
To date, main emphasis has been on the use of wearable sensors that convert physical biometrics such as heart or respiratory rates into electrical signals. For example, ``WEALTHY'' and ``MY Heart'' are EU funded projects that use cotton shirts embedded with sensors to measure respiratory activity, electrocardiograms (ECG), electromyograms (EMG) and body posture \cite{pacelli2006sensing}. The `Sensing Shirt', shown in Fig.~\ref{china} is another example of a shirt capable of monitoring several vital signs including ECG,  posture/activities  in addition to rib cage (RC) and abdominal (AB) respiration, photoplethysmogram (PPG), and SpO$_{2}$ \cite{chinashirt}.  LOBIN is another platform that allows monitoring of physiological parameters and localization of patients within a hospital, using e-textiles and wireless sensor networks \cite{lobin}.

\begin{figure}[h!]
\centering
\includegraphics[angle=-90,width=0.3\textwidth]
{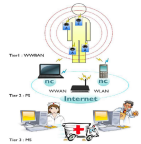}
\footnotesize
\caption{Typical wireless body area network architecture \cite{darwish2011wearable}.}
\label{fig:wireless_architecture}
\end{figure}

Wearable networks are considered multi-stage systems, such as the one shown in Fig.~\ref{fig:wireless_architecture}. The architecture of such a network typically includes in its first stage the nodes of the  wireless body area network. Each node senses, samples, and processes one or more physiological signals. In its second stage, the network architecture includes a personal server application that runs on a personal digital assistant, cell phone or personal computer. The importance of the personal server is that it acts as an interface between the user and the medical server enabling network configuration as well as management features \cite{darwish2011wearable}. The configuration includes registration of the sensor nodes to sort their type and number, initialization to specify the sampling frequency and mode of operation, customization to run user-specific calibration, and security settings communication. Once the wireless wearable network is configured, managing of the network comes next \cite{7425966}. Channel sharing, time-synchronization as well as data retrieval, processing and fusion are amongst the tasks that the personal server application manages. 
The final stage includes  access of the medical server via the Internet. This server typically runs up a service that sets up the communication channel to users, collects reports from the user and integrates the data in the medical record of the user \cite{7392653}. 

%\vspace{-5 pt} 

As an example, such multi-tier systems may be used for analyzing medical data acquired by Wearable Medical Sensors (WMSs) and aiming to assist doctors in disease diagnosis \cite{system}. In this case, data acquired by the WMSs is stored in memory for in and out of clinic monitoring. Robust machine learning ensembles are subsequently implemented, aiming to form a hierarchical multi-tier structure that, in turn, constitues the hierarchical health decision support system. Based on data collected by wearable and/or implantable sensors and given the patient's health history, disease diagnosis can be performed. 

\section{Epidermal Wireless Body Area Networks}\label{sec4}

\begin{figure}[ht!]
  \centering
  \includegraphics[angle=-90,width=0.2\textwidth]{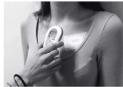}
  \caption{ Epidermal plaster on the human chest wirelessly read by a handheld reader \cite{temp2017}.}
  \label{temp}
\end{figure}

Advancements in material science merged into electronic systems have given way to the development of  epidermal electronics in the medical field. While wearable and textile biomedical sensors are placed on the body and within clothes, respectively, epidermal devices are placed on the skin directly similar to tattoos \cite{rfidTempsensor}. Thin biocompatible membranes are used as substrates for such devices.  Applications of epidermal electronics span a wide range, with temperature monitoring via epidermal UHF RFIDs being a typical example  \cite{rfidTempsensor}, \cite{temp2016}, \cite{temp2017}. 
In \cite{temp2017} temperature sensing and RFID communication were performed using the EM4325 IC and a meander loop antenna. Adhesive copper foil, medical adhesive dressings and biocompatible materials were used to construct the epidermal. With an estimated reading distance of 70 cm, the  aforementioned sensor provides a promising tool for nurses to wirelessly measure the body temperature without contacting the patient as shown in Fig.~\ref{temp}. 
As another example, the feasibility of using Smart Plasters for healing wounds and burns was studied in \cite{hydrogel}. In this case, a loop antenna coupled with an RFID/temperature sensing chip is placed on a biocompatible stretchable substrate made of a polyvinyl alcohol/xyloglucan-based (PVA/XG) hydrogel membrane. The system's response varies with the amount of fluids absorbed by the membrane revealing wound, for example, the presence of wound exudates. Drugs and water can also be delivered to the wound through this hydrogel substrate. 

\begin{figure}[ht!]
  \centering
  \includegraphics[angle=90,width=0.4\textwidth]{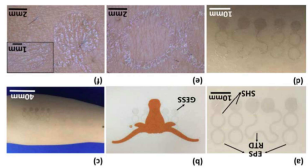}
  \caption{ Graphene-based epidermal electronic sensor (GESS); (a) Fabricated GESS on white background; (b) GESS mounted on tegaderm on top of UT Austin logo showing high transparency. (c) GESS on forearm and (d) close-up view. (e) and (f) Magnified photo of GESS on skin shows intimate conformability to skin \cite{graphene}.}
  \label{graphene}
\end{figure}

In the other case, an electrophysiological sensor (EPS), resistance temperature detector (RTD), and skin hydration sensor (SHS) were all incorporated into a highly transparent graphene- based epidermal sensor system (GESS) \cite{graphene}. With a thickness of around 500nm, this graphene sensor forms one of the thinnest epidermal sensors reported in the literature. As such, the sensor can be placed on the body in a manner similar to a paper tattoo, as shown in Fig. \ref{graphene}.

\section{Implantable Wireless Body Area Networks}\label{sec5}

 Integration of rapid advances in areas such as microelectronics, microfluidics, microsensors and biocompatible materials has recently given rise to implantable biodevices for continuous
medical observation. These sensors act as  event detectors or stimulators that carry out faster, cheaper, and more unobtrusive clinical tasks in comparison to standard methods \cite{7084595}. 

Being used for millions of patients, implantable devices  have already resulted in both improved care and better quality of life for sufferers.   These implanted sensors typically involve  physical, physiological, psychological, cognitive, and behavioral processes, and can reach deep tissue in a reduced response time \cite{cherukuri2003biosec}.
Several implantable sensors for in vivo monitoring are currently being developed \cite{juanola2012market}. An example of a well known implantable identification device is an RFID tag developed by ``VeriChip Corporation''. This device has been intended for implantation in the upper arm where the medical professionals use the serial number emitted by the VeriChip in order to access a patient's medical information in a database called ``VeriMed''. This enables rapid retrieval of vital data even if the person is unconscious or unresponsive during a medical surgery \cite{sullivan2009abc}.  

Other systems enable patients suffering from chronic diseases to live independently. As a result of a joint collaboration in Ohio and California, USA, an implantable kidney has been developed \cite{fissell2007development}. This is considered an alternative to dialysis and transplantation and is based on demonstrated technologies, sound science, and measurable milestones. The developed system actually utilizes efficient membranes and cell-based reactors. Microelectromechanical Systems (MEMS)  have been used to create biocompatible silicon membranes with nano-meter sized pores that can mimic the filtering capacity of the human kidney through cloning \cite{fissell2007development}. In another case, with the parallel development of both  on-chip potentiostat and  signal processing techniques, substantial progress was made towards a wireless implantable glucose/lactate sensing biochip \cite{rahman2009towards}. Implantable bio-MEMS for  in-situ monitoring of blood flow have been reported as well \cite{steeves2007membrane}. The aim was to develop a smart wireless sensing unit for non-invasive early stenosis detection in heart bypass surgery. Chronic blood pressure monitoring devices that are fully implantable are becoming popular as well. Low cost and highly accurate sensors are now possible with the progress of microfabrication technologies. Such implantable sensors allow for continuous assessment of the cardiovascular system conditions, without affecting the patient's daily activities \cite{asimina}.

Even more recently, a fully passive brain implant was developed to acquire   neuropotentials wirelessly \cite{brainimplant}.  The implanted device consisted of electrodes to sense neuropotentials, a mixer, a matching network and a miniaturized antenna. When the device receives a carrier signal from a close by interrogator circuit, the mixer mixes it with the sensed neurosignals before backscattering them to the external device. The signal is then demodulated to retrieve the baseband neuropotentials as low as 20 uV (peak-to-peak). 

Further, the rise of nanotechnology resulted in additional medical research advancements \cite{willander2014nanobiology}. Healthcare is moving quickly towards a future where intelligent medical implants can continuously monitor  body conditions and autonomously respond to changes such as infection by releasing anti-inflammatory agents.  A recent review in WIREs Nanomedicine and Nanobiotechnology, discusses present and prospective implantable sensors incorporating nanostructured carbon allotropes.  Recent progress in nanobiosensors offers technological solutions in the field of glucose monitoring, pregnancy and DNA testing, as well as microRNA detection \cite{juanola2014design}.

Nevertheless, various difficulties need to be addressed when dealing with implantable devices. First and foremost, the device must be biocompatible to avoid unfavorable reactions within the body. The medical device must also provide long-term stability, selectivity, calibration, as well as adequate power in a downscaled and portable device \cite{colomer2013ultra}. Providing power via batteries has been the traditional way, yet they contribute to a major part of the device's size and need to be recharged. Thus energy harvesting technologies, such as harvesting electromagnetic energy, ultra sound, human motion, tissue emotion and heartbeat are  extensively researched to enhance their efficiency \cite{asimina}. Fully passive operation has also been reported, where the in-body device acts as an RFID. The implanted device back-scatters the carrier signal sent by an external interrogator after mixing it with the data it sensed eliminating the need for power storage elements \cite{brainimplant}. 

\section{Powering of Wireless Area Networks}

For wearables and implants, batteries heavily contribute to the size and weight of the sensor. Concurrently, use of batteries implies requirements for frequent recharging and/or battery replacement, both of which are not desirable in this class of applications. Thus, several solutions have been reported for enabling battery free sensors using RFID ICs and energy harvesting techniques, among others. 

For Radio-Frequency (RF) energy harvesting, the density of the power available to the antenna in its surrounding environment is critical and, in fact, determines the powering capabilities. This power depends on the available electromagnetic waves and the effective area of the used antenna, that in turn varies with the wavelength and the actual antenna gain in the specific band \cite{energy}.

Kinetic energy harvesting is one of the most convenient methods for powering wearable sensors by taking advantage of the inherent human
body motion \cite{micro}. Kinetic microgenerators, which may be of two-types, capture forces applied on the miniaturized devices using a mass-spring-damper system. The first resonating type uses the application of the force directly on the device, whereas the non-resonating microgenerator needs a single point of attachment with the body and it exploits the inertial, ambient forces acting on the proof mass \cite{micro}.
 To quantify the kinetic energy, the authors in \cite{kinetic} performed measurements with the energy harvesters placed on the human arm and used a model of the non-resonating topology, namely  Coulomb-Force Parametric Generator (CFPG). In \cite{micro}, the authors enhanced the CFPG model in Simulink to simulate the instantaneous power that can be generated by the CFPG device more accurately.

\begin{figure}[ht!]
  \centering
  \includegraphics[angle=90,width=0.45\textwidth]{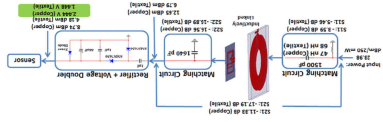}
  \caption{ Measured power flow and two-port measured S-parameters at 11 MHz \cite{piezo}.}
  \label{piezofig}
\end{figure}

A wireless powering model was presented in \cite{piezo} to power an implanted piezoelectric pressure sensor. In this case, the sensor was connected via a circuit to an implanted antenna. This antenna was coupled with an external textile based antenna made up of metalized fabric on ethylene propylene diene monomer (EPDM) cell foam rubber substrate. The material of the latter was chosen so as to make it light weight and comfortable to the patient. Measurements indicated that the received power levels were adequate for feeding the sensor, as depicted in Fig.~\ref{piezofig}.

 As another example of wireless power transfer, the authors of \cite{powering} designed a miniaturized processing circuit that includes an AC-DC converter and a SIDO converter. The circuit combines the three stages of an AC-DC converter (a rectifier, a DC-DC switching converter, and a linear regulator) into one stage hence reducing the size and losses that accumulate per stage. 
Finally, RFID passive, and, thus, batteryless are becoming increasingly popular in wireless sensors. Such tags, which are composed of an antenna and an IC,  receive electromagnetic signals and then transmit them to the reader with the required information \cite{iot}, \cite{chapter}, \cite{temp2016}.

\begin{table}[t!]
\caption{Physical Layer Challenges}\begin{center}
\centering
\small
\label{table:challenges}
\begin{tabular}{|p{110 pt}| p{140 pt}|p{120 pt}|}\hline
\raisebox{10 pt}  

Physical Layer Challenges & Improvement Techniques &  Outcome  \\
\hline
\raisebox{10 pt}

Bandwidth Limitations & The authors in \cite{yuce2009wideband} suggested the use of low data rates and the transmission of multiple pulses per bit.& By implementing the configuration available in \cite{yuce2009wideband}, bandwidth limitations of current narrowband systems is  overcome.  \\\hline
\raisebox{10 pt}

Receiver Complexity& The authors in \cite{thotahewa2014power} suggested the use of Dual Band structure, i.e. using IR-UWB transmitter for the sensor and using narrowband transmitter for feedback at the receiver end.  & Receiver complexity and power consumption is reduced in cross layer design. However, cross layer design is not applicable in the case of low idle time. \\\hline
 \raisebox{10 pt}
 
Power consumption is higher in dynamic conditions.  & The authors in \cite{ho2012ultra} suggested different optimum receiver positions for different sensors. & Transmission power can be reduced by $26$ dB by selecting optimum receiver position for specific sensors. 
\\\hline
\raisebox{10 pt}

Small distance between the transmitter and receiver antennas. & The authors in \cite{see2012experimental} suggested the implementation of different antenna design configurations for different parts of the body & Transmit power can be reduced by $20$ dB or more if optimum type and polarization of antenna are selected for the different body locations. 
\\\hline
\end{tabular}
\end{center}
\end{table}

\section{Challenges  of Wireless Body Area Networks}\label{sec6}
 
\subsection{Physical Layer Challenges}
The physical layer suffers from a number of challenges when it comes to implementing wireless body area networks. The size of the sensor is one issue, which is in many cases due to its antenna. Several methods have been proposed to miniaturize antennas according to their use and substrate they are implemented on. For example \cite{malak} proposed a bouquets like slotted patch on a circular full ground of 3.5 mm radius for implantable sensors, whearas the design in  \cite{amna}  exhibited stacked patches of triangular shape, with one layer having slotted patches causing a size reduction of 60\% compared to previous designs. Circular stacked patches with circular slots were also used to miniaturize the size of the antenna for biomedical telemetry \cite{alaa}. 

Other physical challenges along with suggested improvements are presented in Table \ref{table:challenges}.

\subsection{MAC Layer Challenges}
For health monitoring applications, Quality of Service (QoS) requirements should be explored, particularly for emergency scenarios. The authors in \cite{demirkol2006mac}, proposed a MAC scheme for healthcare applications which merged a preemptive service scheduling into the 802.11.e QoS MAC to provide the highest channel access precedence for medical emergency traffic. Indeed, under emergency conditions, the delivery of data with a reasonable delay should be guaranteed. Accordingly, emergency data prioritization mechanisms should be developed and fairness among different situations should be considered \cite{sadler2006data}. Actually, the prioritization and fairness mechanisms for vital signal monitoring applications are still open research issues. 
\vspace{-5 pt}     
\subsection{Network Layer Challenges}
One of the open research challenges of wireless health care monitoring systems is their capability of reducing the energy consumption of their computing and communication infrastructure. In fact, the convergent traffic inherent in wireless sensor networks may cause a choke effect at the node closer to the base station \cite{fang2010congestion}. Consequently, load balancing routing protocols need to be developed. In addition, congestion avoidance and rate control issues become significant when multimedia traffic is encountered. For better utilization, these techniques should as well be integrated with data compression techniques. 
\vspace{-5 pt} 
 
 \subsection{Transport Layer Challenges} 
 Reliable data delivery is one of the most important requirements of a wireless healthcare network since it deals with life-critical data. Thus, a lost frame or packet of data can cause an emergency situation to be either totally missed or misinterpreted. As a result, a cross layer protocol must be designed in order to ensure reliable delivery of different types of traffic \cite{pereira2007end}.  \vspace{-5 pt} 
 \subsection{Application Layer Challenges}
  Since the application layer is at the top of the stack, it is expected to have a coordinating mission. In this context, the organization of data is critical and requires efficient machine learning algorithms to allow self-learning and autonomous system replacing \cite{6912705}.   
 
\section{Future of Wireless Body Area Network Systems}
Future developments in sensor nodes must produce very powerful, cost effective devices. In this section, we will look into all possibilities of further development in wireless networks in healthcare \cite{6827212}. 

\begin{itemize}

\item From our perspective, several of the aforementioned challenges can be tackled by initially reducing the sensor power consumption in a wireless healthcare network. This may be achieved  through the utilization of code optimization, memory optimization, and  the reliance on less complex data processing techniques. In addition, when the sensor is inactive, a sleeping mode operation should be activated. This allows  higher data rates and enables better time synchronization between the transmitted slots. Further, by reducing wireless data transmission, the protocol overhead will be decreased. This can be achieved through data compression and by transmitting data that is not raw. \\

\vspace{-5 pt} 
\item The development of energy optimization methods is also highly critical. This can be achieved by combining both the link and physical layer functionalities of wireless devices. In turn, this provides longer battery life time, extending the sensor lifetime. \\
\vspace{-5 pt} 
\item The design of high efficiency miniature antennas for sensor nodes is another area of great research interest. The utmost goal is to increase the reliability of transmissions  and minimize interference.
\\
\vspace{-5 pt} 
\item Optimization of each sensor available in the wireless network in accordance with its characteristics is further proposed to be accomplished via a variable sampling rate.
 An adaptive communication protocol is to be used to  accommodate any differences in the system which will make the system power efficient. \\
\vspace{-5 pt} 
\item Design of multiple gateway devices is recommended for interfacing with the existing wireless system. The utmost aim is to ensure continued remote monitoring in a wireless body area network. \\
\vspace{-5 pt} 
\item Sensors must offer flexibility and integration with third party devices and should not operate as standalone systems in a wireless healthcare networks. Wireless body area networks must actually have their own standards in order to collect and store data as well as eliminate any coexistence issues. \\

\vspace{-5 pt} 

\item In highly dynamic environments, wireless sensor networks for healthcare face timing constraints due to severe resource limitations. Several approaches to real time computing like wireless networking protocols, operating systems, middleware services, data management, and theoretical analysis are challenged by wireless sensor networks. Hence, to design time critical systems, different types of systems such as wireless (mesh) sensor networks are used to carry out control processes in real time. \\
\vspace{-5 pt} 
\item  The usage of a cognitive sensor network  is further recommended for acquiring localized and situated information of the sensing environment by the intelligent and autonomical deployment of sensors. Well-known examples for cognitive sensing include both swarm intelligence and quorum sensing. The former is used to study the collective behavior of decentralized, self-organized systems.  The latter has actually gained a lot of interest in the past years as it is an example of bioinspired networking. Specifically, quorum sensing refers to the ability of bacteria to communicate and coordinate behavior via signaling molecules. \\
\vspace{-5 pt} 
\item  Wireless networks should mimic a hierarchical structure that has objectives like scalability, customized services, and energy efficiency. The wireless network should be programmed as a whole rather than programming individual nodes due to the inconsistent behavior of these nodes. In this context, topology control algorithms that provide definite, and practical algorithms is required in order to efficiently measure the network performance and offer idealistic mathematical models. 
\end{itemize}

\vspace{-5 pt} 

\section{Conclusion}
This paper presented a survey of the recent advances in utilizing wireless sensors and body area networks for medical applications with emphasis on wearable, epidermal and implantable technologies. The critical requirements for the design of such sensors were addressed including wearability, security, interoperability, as well as reliable communication. The paper also discussed research challenges related to powering up these medical sensors for various healthcare applications, and addressed future research directions associated with wireless networks for healthcare applications. Fueled by recent advances in both hardware and software, wireless sensor networks is going to result in significant advancements in healthcare practice and research.

%\include{./TeX_files/chapter01}
%\include{./TeX_files/chapter02}
%\include{./TeX_files/chapter03}
%\include{./TeX_files/chapter04}
%\include{./TeX_files/chapter05}
%\backmatter
% bibliography, glossary and index would go here.
\bibliographystyle{IEEEtran}
\bibliography{bibtextry2}

\end{document}